\documentclass[preprint,prb]{revtex4}

\usepackage{amsmath}
\usepackage{amssymb}

\newcommand{\gradv}{\boldsymbol{\nabla}}
\def\v#1{{\bf#1}}

\begin{document}

\title{The Coulomb static gauge}
\author{Jos\'e A. Heras}
\email{herasgomez@gmail.com}
\affiliation{Departamento de F\'\i sica, E. S. F. M., Instituto
Polit\'ecnico Nacional, M\'exico D. F. M\'exico and 
Department of Physics and Astronomy, Louisiana State University, Baton
Rouge, Louisiana 70803-4001, USA}

\begin{abstract}
The existence of gauge conditions involving
second-order derivatives of potentials is not well known in
classical electrodynamics. We introduce one of these gauges,
the Coulomb static gauge, in which the scalar potential is given by the
Coulomb static potential. We obtain an explicit expression for the
associated vector potential and show how the scalar and vector potentials in
this gauge yield the retarded electric and magnetic fields. We
note the close relation between the proposed gauge and the temporal gauge.
\end{abstract}


\maketitle

\section{Introduction}
The topic of electromagnetic gauges has recently received renewed
attention,\cite{1,2,3} 
motivated in part by a paper by Jackson
and Okun,\cite{4} in which the history that led to the
conclusion that potentials in different gauges describe the same retarded
electric and magnetic fields was reviewed. In a subsequent
paper,\cite{1} Jackson developed a simple approach for finding novel
expressions for the vector potential in the Coulomb, velocity, and temporal
gauges and demonstrated how the potentials in these gauges yield the
same retarded electric and magnetic fields. Jackson's approach has been used
to express the retarded fields in terms of the instantaneous fields of a
Galilean-invariant electromagnetic theory\cite{5} and to discuss the
Kirchhoff gauge in which the scalar potential ``propagates" with the
imaginary speed $ic$.\cite{6} The present author\cite{3}
has proposed an alternative method for showing how the equations of
potentials in various gauges yield the same retarded electric and magnetic
fields.

The Lorenz, Coulomb, and Kirchhoff gauges are members of a family
of gauges known generically as the velocity gauge.\cite{1,2,3} The
members of this family are characterized by gauge conditions defined by
first-order derivatives of potentials. For example, the Coulomb gauge condition, $\gradv\cdot\v A=0$. The existence of
gauges defined by gauge conditions involving second-order
derivatives of potentials, which may explicitly involve the sources of the fields, does not seem to have 
been reported before. 

In this paper we introduce the Coulomb static gauge defined by the gauge
condition
\begin{equation}
\label{gauge}
\frac{\partial}{\partial t}(\gradv\cdot \v A)=4\pi c(\rho_s-\rho),
\end{equation}
where $\rho_s=\rho(\v x,t_0)$ is the static charge density, which follows
from evaluating the time-dependent charge density $\rho(\v x,t)$ at an
arbitrarily chosen time $t=t_0$. We use Gaussian units and consider
confined sources in vacuum.\cite{7} The gauge condition (1) is defined by
second-order derivatives of the vector potential and explicitly involves the
charge density, one of the sources of the retarded fields. 

In Sec.~II we show how a vector potential can always be found to satisfy
Eq.~(1). We then apply Jackson's approach\cite{1} to find an explicit
expression for the vector potential in the Coulomb static gauge. In Sec.~III
we calculate the retarded electric and magnetic fields from the explicit
expressions for potentials in the Coulomb static gauge. In Sec.~IV
we show that an equivalent gauge condition for the temporal gauge is given
by 
\begin{equation}
\frac{\partial}{\partial t}(\gradv\cdot \v A)=-4\pi c\rho,
\end{equation}
which can be considered as a special case of Eq.~(1), namely, when $\rho_s=0$. We also emphasize the close relation between the Coulomb
static gauge and the temporal gauge. In Sec.~V we present our conclusions.

\section{The Coulomb static gauge}

It is well known that the electric and magnetic fields $\v E$ and $\v B$ due
to confined sources in vacuum are determined from the scalar and vector
potentials $\Phi$ and $\v A$ according to 
\begin{subequations}
\begin{align}
\v E &= -\gradv\Phi- \frac{1}{c}\frac{\partial\v A}{\partial t},\\
\v B &= \gradv\times\v A.
\end{align}
\end{subequations}
The fields $\v E$ and $\v B$ are invariant under the gauge
transformations
\begin{subequations}
\begin{align}
\Phi'&=\Phi-\frac{1}{c}\frac{\partial \chi}{\partial t},\\
\v A'&= \v A +\gradv\chi,
\end{align}
\end{subequations}
where $\chi(\v x,t)$ is an arbitrary function. The inhomogeneous Maxwell
equations together with Eq.~(3) lead to the coupled equations
\begin{subequations}
\begin{align}
\nabla^2\Phi+\frac 1c \frac{\partial}{\partial t}(\gradv\cdot\v A)&=
-4\pi\rho,\\
\nabla^2{\v A}-\frac{1}{c^2}\frac{\partial^2\v A}{\partial t^2} -
\gradv\Big(\gradv\cdot\v A+\frac 1c\frac {\partial \Phi}{\partial
t}\Big)&=-\frac{4\pi}{c}\v J.
\end{align}
\end{subequations}
The arbitrariness of the gauge function $\chi$ in Eq.~(4) allows us to
choose a gauge condition. For the case of the gauge condition in Eq.~\eqref{gauge}, Eq.~(5) becomes
\begin{subequations}
\begin{align}
\nabla^2\Phi_s&=-4\pi\rho_s,\\
\gradv\times(\gradv\times\v A_s)+\frac{1}{c^2}\frac{\partial^2\v A_s}{\partial
t^2}&=\frac{4\pi}{c}\v J,
\end{align}
\end{subequations}
where $\Phi_{s}$ and $\v A_{s}$ denote the scalar and vector potentials in the Coulomb static gauge.
The novelty of this new gauge is that it is defined by a gauge condition possessing second order derivatives of the vector potential in contrast to the Coulomb gauge which is defined by a gauge condition involving first order derivatives of the vector potential: $\gradv\cdot\v A=0$. The solution of Eq.~(6a) is
given by the Coulomb static potential
\begin{equation}
\Phi_s(\v x,t_0)= \!\int\!d^3x' \frac{1}{R}\rho(\v x',t_0),
\end{equation}
where $R=|\v x-\v x'|$. The potential $\Phi_s(\v x,t_0)$ produces the electrostatic field $\v E(\v x,t_0)=-\gradv\Phi_s(\v x,t_0)$ which satisfies $\gradv\times \v E(\v x,t_0)=0$. It is clear that the term $-\gradv \Phi_s$ does not satisfy the properties of causality and propagation at the speed of light, which are characteristic of the retarded
electric field. Therefore the explicit presence of the static term $-\gradv\Phi_s$ in the expression for the retarded electric field
$\v E =-\gradv\Phi_s - (1/c)\partial\v A_s/\partial t$ seems to suggest some kind of inconsistency. Before discussing this point, we will show that a vector potential can always be found to satisfy Eq.~(1).

Suppose that the original potential $\v A$ satisfies Eq.~(5) but does not
satisfy Eq.~(1), that is, $\partial(\gradv\cdot \v A)/\partial t-4\pi
c(\rho_s-\rho)=g$, where $g=g(\v x,t)\not=0$. We make a gauge
transformation to the potential $\v A'$ and require that it satisfies the
Coulomb static condition 
\begin{equation}
\frac{\partial}{\partial t}(\gradv\cdot \v A')-4\pi c(\rho_s-\rho)=0=
g+\nabla^2\Big(\frac{\partial\chi}{\partial t}\Big).
\end{equation}
Therefore, if a gauge function can be found to satisfy 
\begin{equation}
\nabla^2\Big(\frac{\partial\chi}{\partial t}\Big)=-g,
\end{equation}
the new potential $\v A'$ will satisfy the Coulomb static condition (1) and
Eq.~(6). We note that Eq.~(9) is an instantaneous Poisson equation, which
can be solved by assuming that $\partial\chi/\partial t$ vanishes
sufficiently rapidly at spatial infinity. The associated solution
$\partial\chi/\partial t=f$ can be integrated over time. Therefore, the
existence of a gauge function leading to the Coulomb static condition is
generally guaranteed.

To find the solution of Eq.~(6b) we can apply the approach used in Ref.~\onlinecite{1}. In Eq.~(4) we identify the potentials $\Phi'$ and $\v A'$ with the Coulomb static gauge potentials $\Phi_s$ and $\v A_s$, and the potentials $\Phi$ and $\v A$ with the Lorenz gauge potentials $\Phi_L$ and $\v A_L$,
\begin{subequations}
\begin{align}
\Phi_s&=\Phi_L-\frac{1}{c}\frac{\partial \chi_s}{\partial t},\\
\v A_s&= \v A_L +\gradv\chi_s,
\end{align}
\end{subequations}
where $\chi_s(\v x,t)$ is the gauge function that transforms the potentials
$\Phi_L$ and $\v A_L$ into the potentials $\Phi_s$ and $\v A_s$. From
$\Phi_s$ and $ \Phi_L$ we obtain $\chi_s$ using Eq.~(10a). Hence we can
find $\gradv\chi_s$. Then $\v A_s$ can be obtained from Eq.~(10b) because
$\v A_L$ and $\gradv\chi_s$ are known. 

The Lorenz gauge potentials are
\begin{subequations}
\begin{align}
\Phi_L(\v x,t)&=\!\int\!d^3 x'\frac{1}{R}\rho(\v x',t-R/c),\\
\v A_L(\v x,t)&=\frac{1}{c}\!\int\!d^3 x'\frac{1}{R}\v J(\v
x',t-R/c),
\end{align}
\end{subequations}
From Eqs.~(7), (10a), and (11a) we obtain 
\begin{equation}
\frac{1}{c}\frac{\partial\chi_s(\v x,t)}{\partial t}= 
\!\int\!d^3 x'\frac 1R\rho(\v x',t'=t-R/c)
-\!\int\!d^3 x'\frac 1R\rho(\v x',t_0).
\end{equation}
We integrate both sides with respect to $ct$ to obtain\cite{8}
\begin{equation}
\chi_{s}(\v x, t)= \!\int\!d^3 x'\frac cR\!\int_{t_{0}}^{t-R/c}dt' \rho(\v
x',t') -(t-t_0)\!\int\!d^3 x'\frac cR\rho(\v x',t_0).
\end{equation} 
If we change variables in the first term from $t$ to $\tau= t-t'$, then
Eq.~(13) takes the form 
\begin{equation}
\chi_{s}(\v x, t)= \!\int\!d^3 x'\frac cR\!\int_{R/c}^{t-t_0}d\tau \rho(\v
x',t-\tau) -(t-t_0)\!\int\!d^3 x'\frac cR\rho(\v x',t_0).
\end{equation}
From Eq.~(14) it follows that
we obtain 
\begin{eqnarray}
\gradv\chi_{s}(\v x,t)&=&-\!\int\!d^3 x'\frac{\hat{\v R}}{R}\Big(\rho(\v
x',t-R/c)+\frac{c}{R}\!\int_{R/c}^{t-t_0}d\tau\rho(\v
x',t-\tau)\Big)\nonumber\\
&&{}+c(t-t_0)\!\int\!d^3 x'\frac{\hat{\v
R}}{R^2}\rho(\v x',t_0),
\end{eqnarray}
where $\hat{\v R}=\v R/R$ with $\v R=\v x-\v x'$. If we substitute Eqs.~(11b) and (15)
in Eq.~(10b), then we obtain an explicit expression for the vector
potential in the Coulomb static gauge, 
\begin{eqnarray}
\v A_s(\v x,t)&=&\frac 1c\!\int\!d^3 x'\bigg(\frac{1}{R}\Big[\v J(\v
x',t')-\hat{\v R}c\rho(\v x',t')\Big]_{\rm ret}- \frac{c^2\hat{\v
R}}{R^2}\!\int_{R/c}^{t-t_0}d\tau \rho(\v x',t-\tau) \bigg)\nonumber\\
&&{}+c(t-t_0)\!\int\!d^3 x'\frac{\hat{\v R}}{R^2}\rho(\v x',t_0),
\end{eqnarray}
where the square brackets $[~]_{\rm ret}$ indicate that the enclosed
quantity is to be evaluated at the retarded time $t'= t-R/c$.

Equations (7)
and (16) are explicit expressions for the scalar and vector potentials
in the Coulomb static gauge. The presence of the last term in Eq.~(16) is
crucial to canceling the electrostatic field produced by the scalar field in
the Coulomb static gauge. Also note that the first volume integral in
Eq.~(16) constitutes the vector potential in the temporal
gauge.\cite{1,3} This result anticipates the existence of a close relation
between the Coulomb static gauge and the temporal gauge, which we will
discuss in Sec.~IV.

\section{Retarded fields}

We now show that the fields calculated from the potentials
in the Coulomb static gauge are the familiar retarded electric and magnetic
fields, that is, we will show that 
\begin{subequations}
\begin{align}
\v E &=-\gradv\Phi_s- \frac{1}{c}\frac{\partial\v A_s}{\partial t},\\
\v B &=\gradv\times\v A_s.
\end{align}
\end{subequations}
lead to the retarded form of the fields $\v E$ and $\v B$. We first
calculate $-\gradv\Phi_s$ using Eq.~(7),
\begin{equation}
-\gradv \Phi_s(\v x,t_0)=\!\int\!d^3 x'\frac{\hat{\v R}}{R^2}\rho(\v x',t_0),
\end{equation}
We calculate $- (1/c)\partial\v A_s/\partial t$ using Eq.~(16),
\begin{eqnarray}
-\frac{1}{c}\frac{\partial\v A_s(\v x,t)}{\partial t}&=&-\frac{1}{c^2}\!\int\!
d^3 x'\Big(\frac{1}{R}\Big[\frac{\partial \v J(\v x',t')}{\partial t'}
-c\hat{\v R}\frac{\partial \rho(\v x',t')}{\partial t'} \Big]_{\rm
ret}\nonumber\\ &&{} -\frac{c^2\hat{\v R}}{R^2}\frac{\partial}{\partial
t}\Big\{\!\int_{R/c}^{t-t_0}d\tau\rho(\v x',t-\tau) \Big\}\Big)-\!\int\!d^3 x'\frac{\hat{\v R}}{R^2}\rho(\v x',t_0),
\end{eqnarray}
where we have used the fact that $\partial [~]_{\rm ret}/\partial t= [\partial /\partial t']_{\rm ret}$. After performing the time derivative in the third term on the right-hand side of Eq.~(19) and using $\partial\rho/\partial t=-\partial\rho/\partial
\tau$, we obtain
\begin{equation}
-\frac{1}{c}\frac{\partial\v A_s(\v x,t)}{\partial t}=\!\int\!d^3
x'\Big[\frac{\hat{\v R}}{R^2}\rho(\v x',t')+ \frac{\hat{\v
R}}{Rc}\frac{\partial \rho(\v x',t')}{\partial t'}
-\frac{1}{Rc^2}\frac{\partial \v J(\v x',t')}{\partial t'}\Big]_{\rm
ret}-\!\int\!d^3 x'\frac{\hat{\v R}}{R^2}\rho(\v x',t_0),
\end{equation}
The last term on the right-hand side of Eq.~(20) cancels the
term in Eq.~(18). Therefore, if Eqs.~(18) and (20) are used in Eq.~(17a),
we obtain the retarded electric field in the form given by
Jefimenko\cite{9} 
\begin{equation}
\v E(\v x,t)=\!\int\!d^3 x'\Big[\frac{\hat{\v R}}{R^2}\rho(\v x',t')+
\frac{\hat{\v R}}{Rc}\frac{\partial \rho(\v x',t')}{\partial t'}
-\frac{1}{Rc^2}\frac{\partial \v J(\v x',t')}{\partial t'}\Big]_{\rm ret}.
\end{equation}
Equations (16) and (17b) give the usual expression for the retarded
magnetic field\cite{10}
\begin{equation}
\v B(\v x,t)=\frac{1}{c}\!\int\!d^3x'\frac{1}{R}[\gradv'\times\v J(\v
x',t')]_{\rm ret}.
\end{equation}
Therefore, we have shown that the potentials in the Coulomb static gauge
lead to the retarded fields. 

A recently proposed four-step approach\cite{3} can alternatively be used for showing that the potentials $\Phi_s$ and $\v A_s$ lead to the retarded fields. 
By applying this alternative approach, the reader can obtain the expression 
\begin{equation}
-\frac 1c\frac{\partial \v A_s}{\partial t }=\!\int\!d^3x'\frac{1}{R}\Big[
-\gradv'\rho-\frac{1}{c^2}\frac{\partial \v J}{\partial t' }\Big]_{\rm
ret} + \gradv\Phi_s, \label{23}
\end{equation}
which is equivalent to Eq.~(20).  Equation (23) states that the term $-(1/c)\partial \v A_s/\partial t$ always contains the static
component $\gradv\Phi_s$. Therefore, the explicit presence of the
time-independent term $-\gradv\Phi_s$ in the retarded electric field (17a) is irrelevant
because such a term is always canceled by the component $\gradv\Phi_s $ of
the remaining term $-(1/c)\partial \v A_s/\partial t$. The static field
$-\gradv\Phi_s$ is undetectable and can be interpreted as a
spurious field that exists only mathematically. From
Eqs.~(17a) and (23) we obtain the usual retarded form of the electric field.\cite{10} The  four-step approach leads also to 
the usual retarded form of the magnetic field in Eq.~(22).\cite{10}

\section{The Coulomb static gauge and the temporal gauge}

In this section we will discuss the close relation that exists between the
Coulomb static gauge and the temporal gauge. Traditionally, the temporal
gauge is defined to be one in which the scalar potential is identically
zero,\cite{1}
\begin{equation}
\Phi_T=0.
\end{equation}
In this gauge the electric and magnetic fields are given only by the vector potential
\begin{subequations}
\begin{align}
\v E&=-\frac 1c\frac{\partial \v A_T}{\partial t},\\
\v B&=\gradv\times \v A_T,
\end{align}
\end{subequations}
where we have used the notation $\v A_{T}$ to specify that the vector
potential is in the temporal gauge. Note the
nonexistence of the scalar potential in the temporal gauge, despite a nonzero charge density. This nonexistence of the scalar potential is feasible because the values of the charge density do not necessarily lead to a scalar potential in all
gauges. The existence of a scalar potential generally depends on the adopted
gauge. The retarded values of the charge density always contribute
physically to the electric field, but they do not lead to a scalar potential
in the temporal gauge. 

In the temporal gauge, Eq.~(5) become
\begin{subequations}
\begin{align}
\frac 1c \frac{\partial}{\partial t}(\gradv\cdot\v A_T)&= -4\pi\rho,\\
\gradv\times(\gradv\times\v A_T)+\frac{1}{c^2}\frac{\partial^2\v
A_T}{\partial t^2}&=\frac{4\pi}{c}\v J.
\end{align}
\end{subequations}
Jackson\cite{1} has recently derived an expression for the potential $\v
A_{T}$,
\begin{equation}
\v A_T(\v x,t)=\frac 1c\!\int\!d^3 x'\bigg(\frac{1}{R}[\v J(\v
x',t')-\hat{\v R}c\rho(\v x',t')]_{\rm ret}- \frac{c^2\hat{\v
R}}{R^2}\!\int_{R/c}^{t-t_0}d\tau \rho(\v x',t-\tau) \bigg).
\end{equation}

We can alternatively define the temporal gauge to be one in which the vector
potential satisfies the gauge condition (2). By adopting this condition,
Eq.~(5) become
\begin{subequations}
\begin{align}
\nabla^2\Phi_T&=0,\\
\gradv\times(\gradv\times\v A_T)+\frac{1}{c^2}\frac{\partial^2\v A_T}{\partial
t^2}&=\frac{4\pi}{c}\v J+\frac 1c\gradv\frac{\partial \Phi_T}{\partial t}.
\end{align}
\end{subequations}
Because of the usual assumption that $\Phi_T$ vanishes at infinity, Eq.~(28a) implies that $\Phi_T=0$, which is the traditional form of the temporal gauge condition. Because $\Phi_T=0$, Eq.~(28b) reduces to the usual form given in Eq.~(26b). In other words, if we impose Eq.~(2) as a gauge condition, then the associated potentials are seen to
satisfy Eqs.~(24) and (28b). Therefore, the temporal gauge can
alternatively be defined by either Eq.~(24) or Eq.~(2). 

From Eqs.~(1) and (2) it follows that the temporal gauge can be considered
as a particular case of the Coulomb static gauge, namely, the case
$\rho_s=0$. Furthermore, from Eqs.~(16) and (27) we obtain the connection
between the vector potentials $\v A_s$ and $\v A_T$,
\begin{equation}
\v A_s(\v x,t)=\v A_T(\v x,t)+c(t-t_0)\v E(\v x,t_0),
\end{equation}
where $\v E(\v x,t_0)=-\gradv\Phi_s(\v x,t_0)$. If $\rho_s=0$, then
$\v E(\v x,t_0)=0$ and hence $\v A_s=\v A_T$. If we use Eq.~(29) then we can obtain
the relations 
\begin{subequations}
\begin{align}
-\gradv\Phi_s- \frac{1}{c}\frac{\partial\v A_s}{\partial t}&=-
\frac{1}{c}\frac{\partial\v A_T}{\partial t},\\
\gradv\times \v A_s&=\gradv\times \v A_T, 
\end{align}
\end{subequations}
which show that both gauges yield the same retarded electric and magnetic
fields. 

\section{Conclusion}
In this paper we have introduced the Coulomb static gauge in which the
scalar potential is given by the Coulomb static potential, Eq.~(7). We have
defined the associated gauge condition, Eq.~(1), derived an expression for
the associated vector potential, Eq.~(16), and demonstrated that the
potentials in this new gauge lead to the usual retarded electric and
magnetic fields. We have presented an equivalent gauge condition for the temporal gauge and discussed the
connection of this gauge condition with the Coulomb static gauge condition.
We have also derived the relation between the vector potential in the Coulomb static gauge and the vector potential 
in the temporal gauge, Eq.~(29).

{\bf Acknowledgment}
\vskip10pt

The author is grateful to Professor R. F. O'Connell for the kind hospitality extended to him in the Department of Physics and Astronomy of the Louisiana State University.


\begin{thebibliography}{99}
\bibitem{7}
We define confined (localized) sources as those sources that vanish outside
the surface of a finite region of space. We could also consider sources that
vanish at infinity, that is, sources of order $O|\v x|^{-2-\delta}$ where
$\delta>0$ as $|\v x|$ goes to infinity.
\bibitem{8} Equation (13) contains an additive integration term $\chi_0$
which is a priori a function of $\v x$, but not of time. We can show that
this term is at most a constant. From Eq.~(10) and the fact that the Coulomb
static gauge scalar field is time-independent, we obtain the wave equation
$
\square^2\chi_s=\gradv\cdot\v A_s$. The terms on the right-hand side of Eq.~(13) constitute a particular
integral of this wave equation. Therefore, the term $\chi_0$ is a solution
to the associated homogeneous wave equation. But this solution can consist
only of time-dependent plane waves. A time-independent $\chi_0$,
a solution of Laplace's equation, can at most be a constant if
we demand finiteness at infinity.
\end{thebibliography}
\end{document}